\documentclass[aps,prl,twocolumn,nofootinbib,showpacs]{revtex4}

\def\Lie{\pounds}
\def\dual{{}^\star\!}

\begin{document}
\title{Optimal Choices of Reference for Quasi-local Energy}

\author{Chiang-Mei Chen$^{1,3}$}\email{cmchen@phy.ncu.edu.tw}
\author{Jian-Liang Liu$^{1,3}$}\email{tendauliang@gmail.com}
\author{James M. Nester$^{1,2,3}$}\email{nester@phy.ncu.edu.tw}
\author{Ming-Fan Wu$^{1,3}$}\email{93222036@cc.ncu.edu.tw}

\affiliation{$^1$Department of Physics, National Central University,
Chungli 320, Taiwan} \affiliation{$^2$Graduate Institute of
Astronomy, National Central University, Chungli 320, Taiwan}
\affiliation{$^3$Center for Mathematics and Theoretical Physics,
National Central University, Chungli 320, Taiwan}

\date{\today}



\begin{abstract}
We have proposed a program for determining the reference for the
quasi-local energy defined in the covariant Hamiltonian formalism.
Our program has been tested by applying it to the spherically
symmetric spacetimes. With respect to different observers we found
that the quasi-local energy can be positive, zero, or even negative.
The observer measuring the maximum energy was identified; the
associated energy is positive for both the Schwarzschild and the
Friedmann-Lema\^{i}tre-Robertson-Walker spacetimes.
\end{abstract}

\maketitle

\noindent {\it Introduction.} An outstanding fundamental problem in
general relativity is that there is no proper definition for the
energy density of gravitating systems. (This can be understood as a
consequence of the equivalence principle.)  The modern concept is
that gravitational energy should be non-local, more precisely {\it
quasi-local}, i.e., it should be associated with a closed
two-surface (for a comprehensive review see~\cite{Szabados:2004vb}).
Here we consider one proposal
based on the covariant Hamiltonian formalism~\cite{Nester91} wherein the
quasi-local energy is determined by the Hamiltonian boundary term.
For a specific spacetime displacement vector field on the boundary
of a region (which can be associated with the observer), the
associated quasi-local energy depends not only on the dynamical
values of the fields on the boundary but also on the choice of
reference values for these fields. Thus a principal issue in this
formalism is the proper choice of reference spacetime for a given
observer.

It is generally accepted that the gravitational energy for an
asymptotically flat system should be non-negative and should vanish
only for Minkowski space (see, e.g., \cite{BD68}; for proofs of this
property for GR see ~\cite{Schoen:1979zz, Witten:1981mf}). In view
of this it has been natural to regard these properties as desirable
 for a good quasi-local energy \cite{Szabados:2004vb,Liu:2003bx}.
The idea that a suitable reference should be the one which gives the
minimal energy  then followed quite naturally. We have proposed
using this approach to choose the optimal reference for the
covariant Hamiltonian boundary term.  Here we consider this optimal
choice program for both static and dynamic spherically symmetric
spacetimes (the most important test cases). Specifically, for the
Schwarzschild and the Friedmann-Lema\^{i}tre-Robertson-Walker (FLRW)
spacetimes, we found the resultant quasi-local energies can be
positive, zero, or even negative for different observers. However,
for both cases, there is one observer who would measure the maximum
energy, and for this observer the associated energy is positive.
Furthermore we find that this energy-extremization program (at least
for these spherically symmetric systems)  is equivalent to matching
the geometry at the two-sphere boundary, which provides for a simple
interpretation of the displacement vector.

\medskip
\noindent {\it The Hamiltonian Formulation.} We begin with a brief
review of the covariant Hamiltonian formalism~\cite{Chen:1994qg,
Chen:1998aw, Chang:1998wj, Chen:2000xw}. A {\it first order
Lagrangian 4-form} for a $k$-form field $\varphi$ can be expressed
as ${\cal L} = d \varphi \wedge p - \Lambda(\varphi,p)$. The action
should be invariant under
 local diffeomorphisms, which infinitesimally correspond to a
displacement along some vector field $N$. From Noether's theorem
there is a conserved translational current which can be written as a
3-form linear in the displacement vector plus a total differential:
${\cal H}(N) := \Lie_N \varphi \wedge p - i_N {\cal L} := N^\mu
{\cal H}_\mu + d {\cal B}(N)$. Here ${\cal H}_\mu \equiv - i_\mu
\varphi \wedge \frac{\delta {\cal L}}{\delta \varphi} + \varsigma
\frac{\delta {\cal L}}{\delta p} \wedge i_\mu p$ with $\varsigma :=
(-1)^k$; this identity is a necessary consequence of {\it local}
diffeomorphism invariance (i.e., symmetry for non-constant $N^\mu$).
Consequently ${\cal H}_\mu$ vanishes on shell; hence the value of
the Hamiltonian---the conserved quantity associated with a local
displacement $N$ and a spatial region $\Sigma$---is determined by a
2-surface integral over the region's boundary:
\begin{equation}
E(N, \Sigma) := \int_\Sigma {\cal H}(N) = \oint_{\partial\Sigma}
{\cal B}(N). \label{EN}
\end{equation}
 For {\it any} choice of $N$ this expression defines a conserved
quasi-local quantity.  Different choices of boundary term correspond
to different boundary conditions.

Einstein's gravity theory, general relativity (GR), can be
formulated in several ways. For our purposes the most convenient is
to take the {\it orthonormal coframe} $\vartheta^\mu =
\vartheta^\mu{}_k dx^k$ and the {\it connection one-form}
$\omega^\alpha{}_\beta = \Gamma^{\alpha}{}_{\beta k} dx^k$ as the
geometric potentials. Moreover we take the connection to be {\it a
priori} metric compatible: $Dg_{\alpha\beta} := dg_{\alpha\beta} -
\omega^\gamma{}_\alpha g_{\gamma\beta} - \omega^\gamma{}_\beta
g_{\alpha\gamma} \equiv 0$. Restricted to orthonormal frames where
the metric components are constant, this condition reduces to the
algebraic constraint $\omega^{\alpha\beta} \equiv
\omega^{[\alpha\beta]}$.

We consider the vacuum (source free) case for simplicity. GR can be
obtained from the first order Lagrangian 4-form ${\cal L}_{\rm GR} =
\Omega^{\alpha\beta} \wedge \rho_{\alpha\beta} + D\vartheta^\mu
\wedge \tau_\mu - V^{\alpha\beta} \wedge ( \rho_{\alpha\beta} -
\frac1{2\kappa} \eta_{\alpha\beta} )$, where $\Omega^\alpha{}_\beta
:= d \omega^\alpha{}_\beta + \omega^\alpha{}_\gamma \wedge
\omega^\gamma{}_\beta$ is the {\it curvature} 2-form,
$D\vartheta^\mu := d\vartheta^\mu + \omega^\mu{}_\nu \wedge
\vartheta^\nu$ is the {\it torsion} 2-form, and
$\eta^{\alpha\beta\dots} := \dual (\vartheta^\alpha \wedge
\vartheta^\beta\wedge \cdots)$ is the dual form basis. The 2-forms
$\Omega^{\alpha\beta}$, $V^{\alpha\beta}$ and $\rho_{\alpha\beta}$
are antisymmetric. We take $\kappa := 8\pi G/c^4 = 8 \pi$.
In~\cite{Chen:2005hwa} a ``preferred boundary term'' for GR was
identified:
\begin{equation} \label{expB}
{\cal B}(N) = \frac1{16 \pi} \left( \Delta\omega^\alpha{}_\beta
\wedge \iota_N \eta_\alpha{}^\beta + \bar D_\beta N^\alpha \Delta
\eta_\alpha{}^\beta \right),
\end{equation}
where $\Delta$ indicates  the difference between the dynamic and
reference values and $\bar D_\beta$ is the reference covariant
derivative. The reference values can be determined by pullback from
an embedding of the boundary into a suitable reference space.

\medskip\noindent {\it The Energy-Extremization Program.} Here we
explicitly formulate the extremization program for static
spherically symmetric spacetimes. The Schwarzschild-like metric in
``standard'' spherical coordinates is given by  $ds^2 = - A dt^2 +
A^{-1} dr^2 + r^2 d\Omega_2^2$, where $d\Omega_2^2 = d\theta^2 +
\sin^2\theta d\varphi^2$. However, there are other favorable
coordinate choices for the Schwarzschild metric (e.g.,
Painlev\'e-Gullstrand, Eddington-Finkelstein, Kruskal-Szekeres). In
order to accommodate most well-known coordinates, we consider a more
general version of the Schwarzschild metric via a coordinate
transformation $t = t(u,v), \; r = r(u,v)$; the metric becomes $ds^2
= - ( A t^2_{u} - A^{-1} r^2_{u} ) du^2 + 2 ( A^{-1} r_{u} r_{v} - A
t_{u} t_{v} ) dudv + ( A^{-1} r^2_{v} - A t^2_{v} ) dv^2 + r^2
d\Omega_2^2$. The Minkowski spacetime $d\bar{s}^2 = - dT^2 + dR^2 +
R^2 d\Theta^2 + R^2 \sin^2\Theta d\Phi^2$ is a natural choice for
the reference. However, the essential issue of the reference choice
is the identification between the reference and physical spacetime
coordinates. A legitimate approach is to assume $T = T(u,v), R =
R(u,v), \Theta = \theta, \Phi = \varphi$ and isometrically embed the
two-sphere boundary $S$ at $(t_0, r_0)$ and its neighborhood into
the Minkowski reference such that $R_0 := R(t_0, r_0) = r_0$. Assume
that the displacement vector $N = N^u \partial_u + N^v \partial_v =
N^t
\partial_t + N^r \partial_r = N^T \partial_T + N^R \partial_R$ is future
timelike and the orientation is preserved under diffeomorphisms, i.e.,
$\sqrt{-\alpha} := t_u r_v - t_v r_u > 0$ and $X^{-1} := T_u R_v -
T_v R_u > 0$. The second term of Eq.~(\ref{expB}) vanishes for
spherically symmetric spacetimes;  the energy can then be evaluated:
\begin{eqnarray} \label{expE}
E &=& \frac{r}2 \left( N^u B + N^v C \right) \sqrt{-\alpha} ,
\\
B &=& X T_u + g^{vu} (R_u - 2 r_u) + g^{vv} (R_v - 2 r_v),
\\
C &=& X T_v + g^{uu} (2 r_u - R_u) + g^{uv} (2 r_v - R_v),
\end{eqnarray}
where the subscripts indicate partial differentiations. Note that
the quasi-local energy is evaluated on the boundary two-sphere $S$;
 the variables appearing in Eq.~(\ref{expE}) and in the following
 are also evaluated on $S$. Each choice of the embedding variables
$\{T_u, T_v, R_u, R_v\}$ means a different embedding, hence a
different reference. For any given displacement vector we extremize
the energy with respect to the embedding variables; we get four
equations, but only three are independent:
\begin{eqnarray}
N^u R_u + N^v R_v = N^R = 0, && \label{ETuv}
\\
X^2 T_v (N^u T_u + N^v T_v) - \alpha^{-1} (g_{uv} N^u + g_{vv} N^v)
= 0, && \label{ERu2}
\\
X^2 T_u (N^u T_u + N^v T_v) - \alpha^{-1} (g_{uu} N^u + g_{uv} N^v)
= 0. && \label{ERv2}
\end{eqnarray}
A useful combination $(\ref{ERv2}) \times R_v - (\ref{ERu2}) \times
R_u$ gives
\begin{eqnarray}
X (N^u T_u + N^v T_v) + \alpha^{-1} [ (g_{uv} N^u + g_{vv} N^v) R_u
\nonumber\\
- (g_{uu} N^u + g_{uv} N^v) R_v ] = 0. \label{ERuv}
\end{eqnarray}
From Eq.~(\ref{ETuv}) we  get $R_u = - \frac{N^v}{N^u} R_v$ and $N^T
:= N^u T_u + N^v T_v = \frac{N^u}{X R_v}$; then $R_v$ can be found
from Eq.~(\ref{ERuv}):
\begin{equation}
\frac{N^u}{R_v} - \alpha^{-1} \frac{R_v}{N^u} g(N,N) = 0 \quad
\Rightarrow \quad R_v^2=\frac{\alpha (N^u)^2}{g(N,N)}.
\end{equation}
We require the displacement vector to be future timelike, i.e., $N^T
> 0$ and $N^u > 0$, and the orientation to be preserved under
diffeomorphisms, i.e., the Jacobians are positive. Then $R_v$ should
be positive, and therefore
\begin{equation}
R_v = \sqrt{\frac{\alpha}{g(N,N)}} N^u, \quad R_u = -
\sqrt{\frac{\alpha}{g(N,N)}} N^v. \label{RuRv}
\end{equation}
Now we calculate the energy. Using Eq.~(\ref{RuRv}) we get
\begin{equation}
{\sqrt{-\alpha}}(N^u B + N^v C) = 2 \left( \sqrt{-g(N,N)} - A N^t
\right),
\end{equation}
where the explicit metric is used in the calculation. Choose $N$ to
be unit timelike on the two-sphere, i.e., $-1 = g(N,N) = g_{uu}
(N^u)^2 + 2 g_{uv} N^u N^v + g_{vv} (N^v)^2$, then the quasi-local
energy for any given future timelike displacement vector $N$ reduces
to
\begin{equation}
E = r \left( 1 - A N^t \right), \label{Energy}
\end{equation}
which is independent of the coordinate system. The energy expression
was obtained without knowing the explicit expressions for the
variables $T_u, T_v$. Indeed, we cannot solve for all four embedding
variables, since there are only three independent equations from the
extremization. Thus, this optimal program produces a unique energy
with an equivalent class of references for any given physical
observer. However, it is also reasonable to impose the normalization
condition of the displacement vector in the reference spacetime,
i.e., $\bar{g}(N,N) = -1$. Using this condition and
Eqs.~(\ref{ERu2}, \ref{ERv2}, \ref{RuRv}) we find $T_u = A t_u N^t -
A^{-1} r_u N^r$ and $T_v = A t_v N^t - A^{-1} r_v N^r$. Then the
reference is uniquely determined.

Now we can vary the energy with respect to the displacement vector
to determine the observer who measures the extreme energy. In view
of the constraint $g(N, N) = - A (N^t)^2 + A^{-1} (N^r)^2 = -1$ we
take $\sqrt{A} N^t = \cosh z$ and $N^r/\sqrt{A} = \sinh z$, and then
the energy extremization $\partial E/\partial z = 0$ implies $\cosh
z = 1$ and $\frac{\partial^2 E}{\partial z^2}|_{\frac{\partial
E}{\partial z}=0} \leq 0$. This result means that among all physical
observers the {\it static observer}, i.e., $N =A^{-1/2}
\partial_t$, would measure the {\em maximum} quasi-local energy
\begin{equation}
E = r \left( 1 - \sqrt{A} \right). \label{Emax}
\end{equation}
For Schwarzschild this is a standard value which has been obtained
by many researchers, e.g.,~\cite{Brown:1992br, Chen:1994qg,
Chen:1998aw, Liu:2003bx, Wang:2008jy}.

The energy expression (\ref{Emax}) can also be obtained via a one
step approach of extremizing the energy with respect to the
embedding variables and the displacement vector. We firstly require
the displacement vector to be both the unit timelike Killing vector
of the Minkowski spacetime
\begin{equation}
N = \partial_T \quad \Longrightarrow \quad N^u = X R_v, \quad N^v =
- X R_u, \label{NuNv}
\end{equation}
and a unit timelike vector in the physical spacetime, $-1 = g(N,N)$,
implying
\begin{equation}
X^2 = -({g_{uu} R_v^2 - 2 g_{uv} R_u R_v + g_{vv} R_u^2})^{-1}.
\label{X2}
\end{equation}
With these assumptions, the energy (\ref{expE}) reduces to
\begin{eqnarray}
E &=& \frac{r}{2} X \Bigl[ 1 - \alpha^{-1} X^{-2} + 2 \alpha^{-1}
(g_{uv} r_u R_v - g_{uu} r_v R_v
\nonumber\\
&& - g_{vv} r_u R_u + g_{uv} r_v R_u) \Bigr] \sqrt{-\alpha}.
\end{eqnarray}
Now there are only two embedding variables appearing in the energy
expression. Varying the above energy expression with respect to
these two variables we get
\begin{eqnarray}
(g_{vv} R_u - g_{uv} R_v) (1 + \alpha^{-1} X^{-2}) &&
\nonumber\\
+ 2 R_v (r_u R_v - r_v R_u) &=& 0, \label{ERuA}
\\
(g_{uu} R_v - g_{uv} R_u) (1 + \alpha^{-1} X^{-2}) &&
\nonumber\\
- 2 R_u (r_u R_v - r_v R_u) &=& 0. \label{ERvA}
\end{eqnarray}
The combination $(\ref{ERuA}) \times R_u + (\ref{ERvA}) \times R_v$
gives $X^{-1} = \sqrt{-\alpha}$; then both (\ref{ERuA}) and
(\ref{ERvA}) reduce to the condition $r_u R_v - r_v R_u = 0$.
Together with Eq.~(\ref{X2}) we get $R_u^2 = r_u^2/A$ and $R_v^2 =
r_v^2/A$. We should pick the plus sign for $R_u$ and $R_v$. The
unique energy produced by this program is then $E = r(1 -
\sqrt{A})$, which agrees with what we found above.

\medskip\noindent {\it Energy Measured by Various Observers.} For the
Schwarzschild spacetime $A = 1 - 2m/r$; it is obvious that
Eq.~(\ref{Emax}) can only be valid outside the black hole horizon:
there is no static observer inside the black hole. In order to
discuss the energy inside a black hole, let us examine our energy
formula Eq.~(\ref{Energy}) for a radial geodesic observer in the
Schwarzschild spacetime. For an observer who falls initially with
velocity $v_0$ from a constant distance $r = a$, there are two
different types of orbits: (1) the crash orbit---where ingoing
observers crash directly into the singularity at $r = 0$, while
outgoing observers first shoot out to the turning point
$r_\mathrm{max}$ and then fall back and crash into the singularity;
(2) the crash/escape orbit---where ingoing observers crash, but
outgoing observers can escape to infinity by having a large enough
initial velocity. The displacement vector for such observers is the
unit tangent of the geodesic; then
$N^t = \frac1{1 - 2 m/r} \sqrt{\frac{{1 - 2m/a}}{{1 - v_0^2}}}$,
where $2m < a$ and $0 < r \leq r_\mathrm{max}$. The energy measured
by this observer is
$E = r \left( 1 - \sqrt{\frac{{1 - 2m/a}}{{1 - v^2_0}}} \right)$.
This result agrees with that of the Brown-York quasi-local energy
expression~\cite{Blau:2007wj,Yu:2008ij}. One can see that the energy
decreases as the initial velocity $v_0$ increases. When the initial
velocity $v_0$ is less, equal, or greater than $\sqrt{2m/a}$ (which
is the escape velocity from the Newtonian point of view) the energy
is positive, zero, or negative, respectively. The negative value for
the energy may appear odd, but it can be explained physically.  It
is correlated with the geometric property that the scalar curvature
of the spacelike hypersurface orthogonal to the displacement vector
is (unlike the usual cases) negative.  Note that the ingoing
geodesic observers can measure energy inside the black hole. This
energy is proportional to the radial distance $r$, so it is a smooth
function in the region $0 \leq r \leq r_\mathrm{max}$.

In addition to the static and radial geodesic observers there are
other natural choices---in particular the unit normal of the
constant coordinate time hypersurface in various coordinate systems.
Using this idea in ingoing Eddington-Finkelstein coordinates, $ds^2
= - \frac{du^2}{1 + 2m/r} + \Bigl[ \frac{2 m  du}{r \sqrt{1 + 2m/r}}
+ \sqrt{1 + 2m/r} dv \Bigr]^2 + r^2 d\Omega_2^2$, where $du = dt +
\frac{2m}rA^{-1} dr, dv = dr$, gives $E_{\rm EF} = r [ 1 - (1 +
2m/r)^{-1/2} ] = 2m \bigl( 1 + {2m}/r + \sqrt{1+{2m}/r}
\bigr)^{-1}$. Whereas for ingoing Painlev\'e-Gullstrand coordinates,
$ds^2 = - du^2 + \bigl( \sqrt{2m/r} du + dv \bigr)^2 + r^2
d\Omega_2^2$, where $du = dt + A^{-1} \sqrt{2m/r} dr, dv = dr$, the
energy {\em vanishes}. All these outcomes are smoothly dependent on
$r$.

\medskip\noindent {\it The Geometrical Meaning.} The
proposed optimal program can be interpreted as an adapted coordinate
choice for any given displacement vector, i.e., observer. The
adapted coordinates, denoted as $\{ u', v' \}$, have the associated
coframes
\begin{equation}
\vartheta^0 = a^0_{u'} du', \qquad \vartheta^1 = a^1_{u'} du' +
a^1_{v'} dv',
\end{equation}
and the displacement vector is the unit timelike vector, namely $N =
e_0$; its components can be expressed in terms of $(a^0_{u'},
a^1_{u'}, a^1_{v'})$. The embedding variables are determined by
requiring on $S$
\begin{eqnarray}
\vartheta^0 = a^0_{u'} du' &=& dT = T_{u'} du' + T_{v'} dv',
\nonumber\\
\vartheta^1 = a^1_{u'} du' + a^1_{v'} dv' &=& dR = R_{u'} du' +
R_{v'} dv'. \label{isomatch}
\end{eqnarray}
This identification leads directly to (\ref{Energy}). Thus, the
extremization program, in these adapted coordinates, yields an
isometric embedding on the two-sphere boundary in a way such that
$N$ is the unit timelike vector in both the physical and Minkowski
reference spacetimes.

\medskip\noindent {\it Dynamic Spherically Symmetric Spacetimes.} The
optimal reference choice program can likewise be applied to dynamic
spherically symmetric spacetimes, the FLRW cosmological models. With
a calculation similar to that above, we found that the optimal
energy for a given displacement vector is
\begin{equation}
E = a r \Bigl( 1 - \sqrt{1 - k r^2} N^t - \frac{a \dot{a} r}{\sqrt{1
- k r^2}} N^r \Bigr),\label{flrwE}
\end{equation}
where $\dot{a} = da/dt$ and $k = -1, 0, 1$ is the sign of the
spatial curvature. An obvious choice  is the {\it comoving}
observer, $N =
\partial_t$, then the energy is $E = a r \left( 1 -
\sqrt{1 - k r^2} \right)$---a value which has been found
previously~\cite{Nester:2008xd}---which is {\em negative} for
negative curvature. On the other hand, varying the
energy~(\ref{flrwE}) with respect to the displacement vector we find
the maximum energy
\begin{equation}
E = \frac{a r^3 (k + \dot{a}^2)}{1 + \sqrt{1 - k r^2 - \dot{a}^2
r^2}}, \label{E1FLRW}
\end{equation}
(which again does not exist for all observer locations $r$).  The
associated displacement vector is
\begin{equation}
N = \frac{\sqrt{1 - k r^2}}{\sqrt{1 - k r^2 - \dot{a}^2 r^2}}
\partial_t - \frac{\dot{a} r}{a} \frac{\sqrt{1 - k r^2}}{\sqrt{1 - k
r^2 - \dot{a}^2 r^2}} \partial_r,
\end{equation}
which is just the unit dual mean curvature
vector~\cite{Tung:2007vq}. For a dust model with matter energy
density $\rho$, by imposing the Friedmann equation, $k + \dot{a}^2 =
(8 \pi/3) \rho a^2$, the energy expression (\ref{E1FLRW}) becomes
\begin{equation}
E = \frac{(8\pi/3) \rho a^3 r^3}{1 + \sqrt{1 - (8\pi/3) \rho a^2
r^2}} \geq 0.
\end{equation}
For small $r$ the proper matter interior limit required by the
equivalence principle is evident. While we find that the observer
dependent quasi-local energy can be {\it negative}, for the
particular observer who measures the maximum value, the quasi-local
energy is {\em positive}.

\medskip\noindent {\it Conclusion.} The covariant Hamiltonian
quasi-local energy expression (\ref{expB}) has certain nice
properties,~\cite{Chen:1994qg, Chen:1998aw, Chang:1998wj,
Chen:2000xw, Chen:2005hwa}, but it suffers from two ambiguities:
which displacement vector and which reference? We have proposed
isometrically embedding the two-sphere boundary and its neighborhood
in the dynamic spacetime into the Minkowski reference and then
extremizing the energy to determine the embedding variables. Here we
have discussed the application of this program to static and dynamic
spherically symmetric spacetimes where we obtain, for any given
future timelike displacement, a unique quasi-local energy which can
be positive, zero, or even negative. When we further vary the energy
with respect to the displacement vector we can identify a special
observer who measures the maximum---and always positive---energy.

Moreover, the optimal program is, at least in this spherical case,
actually equivalent to an adapted coordinate choice for each
observer. In such coordinates, our program not only isometrically
matches the geometry near the two-sphere boundary,  but also
identifies the displacement vector as the unit timelike vector
orthogonal to the constant coordinate time hypersurface, and, at the
same time, as the unit timelike Killing vector of the Minkowski
reference. This observation lends further support to the proposed
optimal program.

We believe that this spherical case is the main test case; it shows
that our program has promise as a universal approach for determining
the reference needed for the covariant Hamiltonian boundary term
quasi-local energy for general spacetimes.

This work was supported by the National
Science Council of the R.O.C. under the grants NSC-98-2112-M-008-008
(JMN) and NSC 96-2112-M-008-006-MY3 (CMC) and in part by the
National Center of Theoretical Sciences (NCTS).


\end{document}